\documentclass[english,11pt]{article}
\usepackage{graphicx,rotating}
\newcommand{\moy}[1]{\left\langle #1 \right\rangle}

\begin{document}

\title{Precise validation of neutron cross section data bases\\
 using a lead slowing down spectrometer \\
 and simulation from 0.1~eV to 40~keV.\\
  Methodology and data on thin and thick targets\\
  and data bases adjustement}
\author{L.Perrot, A.Billebaud, R.Brissot, A.Giorni, D.Heuer, \\
J.-M.Loiseaux, O.M\'eplan, and J.-B.Viano\\
Institut des Sciences Nucl\'eaires, IN2P3-CNRS/UJF\\ 53, av. des Martyrs, 38026 Grenoble Cedex France}
\date{\today}
\maketitle
\abstract{Research on accelerator driven systems (ADS), related new fuels and their ability for
nuclear waste incineration has led to a revival of interest in nuclear cross sections of many nuclides over a large energy range.
Discrepancies observed between nuclear data bases require new measurements in several cases. A complete measurement of such cross
sections including resonance resolution consists of an extensive beam time experiment associated to a long process of
analysis ans validation. With a slowing
down spectrometer associated to a pulsed neutron source, it is possible to determine good cross section profile in an energy range
from 0.1~eV to 40~keV. These measurements performed at ISN (Grenoble) with neutron source GENEPI requires only small quantities of
matter (as small as 0.1~g) and about one day of beam per target.\\
We present measured cross section profiles and an experimental study of self-shielding effect. A CeF$_3$ scintillator coupled with a
photomultiplier detects gamma rays from neutron capture in the studied target. The neutron flux is also measured with a $^{233}$U
fission detector and a $^3$He detector placed symmetrically to the PM in relation to the neutron source. Absolute flux values are
given by activation of Au and W foils. Test of slowing down properties is done with a precision of 1\% The cross section profiles 
can then be deduced from the target capture rate and are compared
with very detailed MCNP code simulations, which reproduce the experimental set-up and provide also capture rates and neutron flux. A
good agreement between experimental and simulated profiles for well-known cross sections like Au for different thicknesses is found in
our energy range, and therefore validates the method and the taking into account of self-shielding.\\
The method is then applied to $^{232}$Th which is of main interest for new fuel cycle studies, and is complementary to higher energy
measurements made by Karamanis et al. \cite{karamanis}. Results obtained for three target thicknesses will be compared with the
simulations based on different data bases. Special attention will be paid to unresolved resonances region (E$>$ 100~eV).}

\newpage
\tableofcontents

\newpage
\section*{Introduction}

Waste incineration, durable nuclear energy production with  new fuels, new coolant and structural materials, some
discrepancies or lack of precision of the nuclear data bases bring clearly new needs in terms of experimental
validation of cross-sections and
associated deviation.

Using a Lead slowing down spectrometer coupled to a specially designed pulsed neutron generator and new types of
detectors with modern data acquisition system, it will be shown that very precise experiment performed on
(n,$\gamma$) or (n,f) cross-section leads to validation of cross-section energy profile within 5\% error.  In
addition, with the same experimental equipment, we present also a precise check of the Lead medium transport
properties (slowing from 14~MeV down to 0.2~eV, elastic and capture cross-section between 40~keV and 0.2~eV).

On the other hand it is demonstrated that the Monte-Carlo Code MCNP is able to predict precisely reaction rates.

The very careful analysis of both experimental and simulation results should allow a precise validation or 
invalidation of neutron cross section over a large energy range.

The present paper is going to show a few significant examples where clearly some neutron cross sections are 
inadequate to describe reaction rate, while some others do it with a relative precision of 5$\%$ over more 
than 5 orders of magnitude in energy (from 40~keV to 0.2~eV).

In addition, experimental and simulation data have been precisely obtained for thin and thick targets, which 
demonstrates the validity of cross section data for describing important self-shielding effects, in large resonances
as well as in the unresolved resonance region.

The general methodology is described in some details and tends to show that this method can be very useful in 
test and validation of data bases for neutron cross sections and is able to propose the best choice
between data bases as long as an energy dependent adjustement of these is needed.

\section{The lead slowing down spectrometer (LSDS) and principal properties}
\label{section1}

\subsection{Physics}
\label{phy_spectro}

In a lead medium, the average neutron velocity change between two
elastic collisions is approximately constant:

\begin{equation}
\label{Eq:velo}
\frac{\Delta v}{v}=\alpha =\frac{m_{\rm Pb}\times m_{\rm n}}{(m_{\rm Pb}+m_{\rm n})^{2}}
\end{equation}

\noindent \( m_{\rm Pb} \) and \( m_{\rm n} \) are respectively
the Lead nucleus and neutron masses~; this leads to a correlation
between the time \( t \) at which a neutron is observed and its velocity
\( v \) \cite{berg55, berg58, rubb95, saw74, tarc}:

\begin{equation}
\label{Eq:tv}
t=\frac{(1-\alpha )\lambda _{s}}{\alpha }\left( \frac{1}{v}-\frac{1}{v_{0}}\right) 
\end{equation}

\noindent where the mean free path \( \lambda _{s}\approx 2.76 \)~cm
in the Lead is approximately constant and \( v_{0} \) is the initial
neutron velocity. This correlation could be rewritten in a slightly
different way, using the time and the neutron kinetic energy \( E \):

\begin{equation}
\label{Eq:E_t}
E=\frac{K}{(t+t_{0})^{2}}
\end{equation}

\noindent where \( K=\frac{m_{\rm n}\lambda _{s}^{2}(1-\alpha )^{2}}{2\alpha ^{2}} \)
and \( t_{0}=\frac{(1-\alpha )\lambda _{s}}{\alpha }\sqrt{\frac{m_{\rm n }}{2E_{0}}} \) \footnote{%
Sometimes, this relation is expressed with the average lethargy change
which is \( \xi \sim 2\frac{\alpha }{1-\alpha } \) 
}. The time \( t_{0} \) could be understood as a correction to the
fact that neutron is not created at infinite energy but at \( E_{0} \)
(14~MeV for the tritium target). In practice, inelastic scattering
on lead is predominant for high energy neutrons, thus this parameter
will be a phenomenological constant that we have determined by a detailed
Monte Carlo simulation.

The relation (\ref{Eq:E_t}) will be very help-full to determine the
energy of the detected neutron, measuring the laps of time between
the start of the beam pulse and the time of neutron interaction in
the detector.

The energy--time relation is not a one-to-one function but a mean
value of a correlation. This correlation has been studied in detail
in the TARC experiment \cite{tarc} and the same kind of analysis has
been done here.

\subsection{Description of the facility}

The slowing down time spectrometer is a cubic assembly of 46.5 tons of pure lead
(99.99\%): it consists of 8 blocks of $80\times 80\times 80$~cm$^3$. A central
channel permit the insertion of the glove finger of the accelerator ended by the
target (described in the following section) up to the middle of the
spectrometer (mechanical sketch can be seen in figure~\ref{genepi}). Each block has two channels ($10\times 10$~cm$^2$ in section)
parallel to the beam axis, which are used both for handling the block and
detector insertion. With lead bricks they can be reduced to 5$\times$5~cm$^2$ in
the case of detector insertion. Pure lead was chosen to ensure that impurities, principally
Silver, Bismuth, Cadmium, Copper, Antimony and Tellurium are less than 5 ppm 
and have thus negligible effect on the neutron flux. The lead block is shielded
with a Cadmium foil to capture neutrons that escape from the block and are
backscattered by concrete walls, and may perturb the time-energy correlation. 
A simulation shows that the principal contribution of backscattered neutrons comes from the support 
of the Lead block. \\
 
\subsection{The neutron source}

The neutron source coupled with the spectrometer is a pulsed source especially
designed for neutronic experiments carried out either with LSDS or in the nuclear reactor MASURCA at
the CEA Cadarache Center (France). Operating
in a sub-critical reactor or in a LSDS, the pulse intensity must as big as possible. The
GENEPI ({\bf GE}nerateur de {\bf NE}utron {\bf P}uls\'e {\bf I}ntense) consists
in a duoplasmatron source producing fast deuteron pulses of less than 1~$\mu$s.
This pulse duration corresponds to few neutron lifetimes in subcritical reactor and to the slowing down time
necessary for a 14~MeV neutron to reach 100~keV. 
The repetition rate of the source can vary from a few Hz up to 5~kHz. The
deuterons are accelerated to the maximal energy of 250~keV, with a maximum peak
intensity of 50~mA. They are then focused through a five meter long tube (glove
finger) onto a deuterium or tritium  Titanium target. The nuclear reactions
D(d,n)$^3$He or T(d,n)$^4$He produce neutrons with energy 2.67~MeV and 14~MeV
respectively. With a tritium target the source can produce typically 5.10$^6$
neutrons per pulse in 4$\pi$.\\

\section{Experimental and simulation methodology}
\label{section2}

\subsection{The observables}
\label{observables}

	\subsubsection{E-t correlation and absolute neutron flux energy variation measurements}

The E-t correlation and absolute flux energy variation measurements allow to have precise check of $\sigma_{el}(E)$,
$\sigma_{(n,\gamma)}(E)$ in Lead, as well as energy variation of $\Phi$(E$<$~40keV) per source neutron.
The 14~MeV neutrons delivered by the generator are slowed down below 800~keV mainly by (n,2n) reactions
followed by inelastic scattering on different Lead isotopes in less than a few 10$^{-7}$ second. 


Then the only slowing down process is the elastic scattering. For
isotropic elastic scattering, the relation (\ref{Eq:tv}) can be rewritten
(assuming initial energy of the neutron is \( E_{0} \), and energy
at time \( t \) is \( E \))

\begin{equation}
t=\frac{1-\alpha }{\alpha }\lambda _{s}(E)\left( \sqrt{\frac{m_{\rm{n}}}{2E}}-\sqrt{\frac{m_{\rm{n}}}{2E_{0}}}\right) 
\label{t_ral}
\end{equation}

where \( \lambda _{s}(E)=\frac{1}{N_{\rm{at}/\rm{cm}^{3}}^{\rm{Pb}}\, \sigma ^{\rm{Pb}}_{\rm{el}}(E)} \).


From relation (\ref{t_ral}), it appears that the slowing down time is directly related to elastic cross section
absolute value behaviour between E$_0$ and E$_f$. Measuring this slowing down time between 40~keV and 0.2~eV povides a
good test of $\sigma_{el}$ behaviour through relation (\ref{t_ral}). We may also point out  that if $\moy{\sigma_{el}}$ has only a
smooth
energy variation, as it is the case of lead, a large part of the slowing down time (68\%) is spent to go from $E=10E_f$ to $E_f$,
which means that the mean level of $\moy{\sigma_{el}}$ can be tested from 2~keV down to 0.2~eV.
This slowing down time can be measured with a precision of 2\%, giving a validation of $\sigma_{el}$ in the few percent range, at
least for $t> 10~\mu s$  ($E< 1.7~keV$).\\
Moreover the time dispersion observed from well known capture resonances for
going from E$_0$ to E$_f$ is also a good test of the simulation code.\\
We will show in the section \ref{e_t} how this precise energy time correlation is sensitive to elastic scattering.\\

The neutron flux energy variation can be precisely measured via the time variation of reaction rate
$\phi(t,E)\sigma(E)$ over 0.2~eV up to 40~keV if $\sigma(E)$ is well determined.\\
To measure $\phi(t(E))$, we used two kind of detectors (described below) based on two well known cross section
($\sigma_{(n,p)}^{^3He}$ and $\sigma_{(n,\gamma)}^{Au}$) in the above energy range.\\
Such precise measurement is sensitive to $\sigma^{Pb}_{el}(E)$ and $\sigma^{Pb}_{(n,\gamma)}(E)$ behaviour. Namely in a
given range of energy an over-evaluation of $\sigma_{el}$ will be observed as a flux under-predicted as shown in \cite{nif}. The
(n,$\gamma$) absorption in Lead is also tested through $\phi(E)$ variation since relation (\ref{t_ral}) can be used
to evaluate the flux attenuation in the case of $\sigma_{el}= Cte$ and $\sigma_{(n,\gamma)}(E)=\frac{A}{v}$. Within
these assumption, $\phi(t(E))$ is attenuated through (n,$\gamma$) process by an attenuation factor which writes as
$e^{-\beta t}$ where $\beta=f(A,\sigma_{el})$. The associated precision on the Lead (n,$\gamma$) cross section absolute value will
be estimated in section~\ref{flux_measure}.\\
	
Absolute flux measurement and absolute calibration of the neutron source have been performed. The ex\-pe\-ri\-men\-tal method will be described in
sections~\ref{neutron_source} and \ref{u3_detector}. 
The measurement of absolute flux in the energy range 0.2~eV to 40~keV for a measured number of source neutrons gives the
quantity $\Phi$(E$<$~40keV) per 14~MeV source neutron which can be compared directly with simulation data. These absolute measurements
allow the test of the transfer function for the neutron from 14~MeV down to 40~keV, transfer function which is sensitive to elastic
or inelastic scattering,(n,2n), and (n,$\gamma$) reactions in Lead.\\

	\subsubsection{(n,$\gamma$) reaction rate for samples}

Energy correlation in a lead block is characterised by an approximate equation (\ref{k_om}) which writes as
\begin{equation}
E(keV)=\frac{K(keV/\mu s^2)}{(t(\mu s)+t_0(\mu s))^2}
\end{equation}
and a dispersion of time for a given energy of about 20\%.\\
When measuring at time t, reaction rate $\phi\sigma_{(n,\gamma)}$ on a small sample of material in the LSDS, the measurement is
sensitive to the actual mean flux in the sample volume. The mean flux is first sensitive to the slowing down process
in Lead up to the time t, and also to the average flux in the sample. Even with relatively thin foils, due to very
large (n,$\gamma$) cross sections in resonances, self shielding effect has to be taken into account. In order to test
precisely the description of these effects by the simulation codes we have made measurements on samples of
different thicknesses (from 100~$\mu$m up to a few millimeters). Finally, if the energy variation of flux is well
described and precisely validated by flux measurements, then the $\phi\sigma_{(n,\gamma)}$ reactions will permit to
validate cross section data bases, or to indicate the amplitude of deviations in some energy range.\\

\subsection{The associated detectors and acquisition}

	\subsubsection{Neutron source monitoring} 
\label{neutron_source}

The monitoring of the source was made by detection of the associated particle of the neutron producing reaction:
alpha in the case of T(d,n)$^{4}$He reaction, proton in the case of D(d,p)T reaction which occurs about as often as
the D(d,n)$^{3}$He reaction. For 0$^\circ$ neutrons the associated particle is emitted in the opposite direction and
goes upstream the incident deuteron beam, suffering focalization in the glove finger and being bent in the GENEPI
magnet chamber (having both the same mangnetic rigidity) to finally impige one of the two silicon detectors placed side by 
side in the vacuum of this chamber. One 
detects alpha and proton while the other, covered with a 10$\mu$m Aluminium foil, detects only protons. The
detection of alpha relatively to protons gives the 14~MeV neutron source relatively to the 2.5~MeV source which appears
because of deuteron ion implantation in the target. The ratio of these two sources was typically equal to 50.\\ 

	\subsubsection{$^{3}$He detector for relative normalization} 

A classical $^3$He proportional counter has been used for neutron flux monitoring. This detector belongs to a serie
of counters especially designed for neutron flux spectroscopy in a fast sub-critical reactor. They have to fit special 
geometrical and efficiency requirements 
and therefore are  built and filled at ISN. The gas filling of the detector used during this experiment is 70~mbars 
of $^3$He, 6 bars of Argon and a few 
mbars of CO$_2$ (as quenching gas). Its detection effective zone is a cylinder of 6~cm long and 2~cm in diameter. 
This counter has been placed in a channel symmetrical to the capture rate counting device in relation to the beam
axis, and has remained at 
the same place during all the measurements. Its counting rate above a constant threshold was used for neutron
spectrum normalization.\\
Time spectra have also been performed with this detector to obtain neutron flux as it will be shown in section~\ref{abs_nflux}.\\

	\subsubsection{$^{233}$U detector and activation for neutron flux measurements} 
\label{u3_detector}
A $^{233}$U target facing a silicium detector placed in a small Lead box was used to measure the neutron flux. This target consists
in a 1~cm$^2$ deposit of 200~$\mu$g/cm$^2$ of $^{233}$U on a 100~$\mu$m Aluminium foil. It is collimated to 0.85~cm  in diameter. 
The fission fragments were detected in the silicium as well as the alpha particle (beam off) for a self-calibration of the
detector. Flux has been measured at a symmetrical position to the (n,$\gamma$) measurements in relation to the beam axis.\\
For absolute flux measurements Nickel targets have been irradiated at different places in the Lead block. With the Tritium neutron
source the two reactions exploited are $^{58}$Ni(n,2n)$^{57}$Ni, and $^{58}$Ni(n,np)$^{57}$Co. These reactions having a
threshold around 13~MeV, they allowed to study the propagation of the 14~MeV source neutrons in the block.\\

	\subsubsection{(n,$\gamma$) detector} 

Capture reaction rates have been measured with a small photomultiplier Philips XP1911 of 18.6~mm in diameter. A teflon embase has
been especially built for these measurements in order to avoid Hydrogen. Decoupling capacitors have been modified to allow a
decay constant long enough compared to the time between two pulses. A CeF$_3$ scintillator with a 30~ns time constant was coupled to
the PM: this material has been chosen for its low capture cross section. The only contribution of Hydrogen is coming from the
adhesive surrounding the PM: despite it is supposed to be Hydrogen free there is a residual proportion. 
The use of a Lead box where the PM and the target are embeded ensured the reproducibility of the capture measurements.\par
A LED was coupled to the scintillator in order to measure the PM gain variation. This variation has been measured less than
4\% in the energy range 0.1~eV to 50~keV, insuring the good reproducibility of the measurements.\par
We have made a systematic determination of the average position of the collected gamma pulse distribution for each sample. 
For each element we obtained this position as a function of neutron energy. We have seen that whatever the sample studied, this
position varies only within 5\% in the energy range 0.1~eV-30~keV. In this energy range neutronic capture happens mainly in the
L=0 angular momentum state, which means it seems reasonable to think that the detector efficiency does not depend on neutron
energy. We will see in section~\ref{au_validation} and \ref{application} the good agreement between simulated and experimental reaction rates in the
0.1~eV and 30~keV energy range. Apart from the discrepancies due to cross section description, we observe an agreement better
than 5\%. This indicates strongly that the gamma detection efficiency does not vary more than 5\% in this range and with the
chosen experimental conditions.\\

	\subsubsection{Acquisition}

Detector signals are coded with an ADC designed at ISN and providing also time referenced to the neutron pulse for each
event. The sampling time, which can be as low as 25~ns, is here chosen to 100~ns. To avoid dead time mainly due to
analogic coding (6.2~$\mu$s), an electronic veto was used to suppress counting during the intense source pulse. With the same aim beam
intensity has been adjusted to have less than 0.1 count per pulse measured by the PM over the first 10~$\mu$s after the pulse, this region
defining the high energy limit of this measurement.\par
	
\subsection{Simulation}

Simulations are made with the MCNP (Monte Carlo N-Particles) code. This Monte Carlo transport code has been developped at
Los Alamos Laboratory by John Hendricks, Judith Briesmeister et al. \cite{mcnp}.We use here MCNP-4B version, exploiting
only the neutron transport module. With this code it is possible to make a very precise description of the experimental
set-up geometry: concrete experimental hall, GENEPI glove finger, lead block, detectors, as it can be seen in
figure~\ref{geometry}. The neutron source due to T(d,n)$^4$He reactions has been described using data from \cite{ndt}. \\
Simulations depend on two main effects. First, materials composing each item given by constructors, for instance lead
impurities which exact proportions are not quite well known, can have high absorption resonances on neutron flux (see
section~\ref{background}). Hydrogen proportion in adhesive and glue surrounding the PM, and thus very close to the target
studied, is also of major importance. Elastic collision of a neutron on hydrogen leads to a mean energy loss of half of its energy and therefore the
neutron does not follow the time energy correlation law any more. It brings an increase the reaction rate for short detection times and
induces a filling up of the spectra between the resonances (figure~\ref{hydrogen}). Nuclear Magnetic Resonance
measurements have been necessary to know precisely the hydrogen proportion in adhesive and modify the associated material description
in the simulations. \\
The other main contribution in simulations is the choice of the data base used for the  materials the most significant on
neutron transport. We will see in section~\ref{section3} that we have settled on JENDL3.2 data base for Lead.\\
Then for a fixed geometry and composition the only free parameter for the targets of interest is the data base used.\\
Calculated reaction rates versus time are made up to 3~ms, and their time bining is chosen the same than the experimental
one, i.e. 100~ns.\\

\subsection{Data processing, background substraction and comparison methodology for (n,$\gamma$)}
\label{background}

A specific procedure has been developped to extract capture rates measured  by the photomultiplier. This method is
applied to all the targets, radioactive or not, thick or thin. Constant levels are first of all substracted from the background and the sample time spectra.
These levels are determined by fitting the spectra at times greater than 4~ms. They are associated to the activation of
surrounding materials such as scintillator, PM, Lead, impurities, to the possible activation of the target itself, and
sometimes to the intrinsic activity of the target. Secondly, corrections must be applied to the background spectrum to
take into account effects appearing only when the sample to be studied is in front of the scintillator. They are of two
types: $\gamma$-rays emitted by various materials contributing to the background can be damped, and the neutron flux can
be also locally damped because of the presence of the target. Then the corrected background can be subtracted to the sample 
spectrum and the normalized capture rate spectrum is obtained.\\

Simulated reaction rates as a function of time calculated for $\delta$ neutron injections are folded by the GENEPI pulse shape. They are normalized 
to the same number of neutrons than the experimental spectra (themselves normalized by the counting of the $^3$He monitor).\\

For the background spectrum which is of first importance, it has been necessary to investigate every spectrum structures and the
associated component proportions surrounding the target. This analysis has led to a background reconstruction whose results are
shown in table~\ref{bdf1} and in figure~\ref{bdf2}.\\

By applying the E-t relation~(\ref{t_ral}) to the simulated and experimental spectra, reaction rates per monitor
neutron versus energy are obtained. This presentation will be adopted to simplify the understanding of the spectra. \\

 \par

\section{E-T correlation and flux measurements}
\label{section3}

	\subsection{The time-energy relation in the lead block}
\label{e_t}

As we have seen in section~\ref{phy_spectro} and \ref{observables}, the neutron slowing down time in the lead is related to its
kinetic energy through the relation $E=\frac{K}{(T+t_0)^2}$. We have also shown that the parameter K was depending on the
elastic cross section of Lead. In figure~\ref{param_k} we show the variation of this parameter deduced from the neutron flux
simulation versus energy and time, respectively by use of ENDF/B-VI and JENDL3.2 base for Lead. We can deduce this parameter
experimentally for a precise resonance energy of several samples like Au, In, and Mo.\\
First of all we can notice that K is not independant on the neutron energy. This parameter is sensitive to the diffusion cross
section. From the relation~(\ref{k_om}) we can deduce $\frac{\Delta K}{K}=\frac{2\Delta\sigma_{el}}{\sigma_{el}}$. At 100~eV, a
difference of 1.2\% exists between elastic cross sections of the two data bases, and it leads to a 2.4\% differnce on the
associated K values. We observe also below 1~eV that K values increase whatever the data base chosen. This increase is caused by
two reasons. First of all, the part of neutrons escaped from the Lead block but going back into it becomes non negligeable below 5~eV: we can
notice that, though they are correlated in time and energy, they need more time to reach the kinetic energy they might have
without escaping from the block. Having a slowing down time larger, the K value is increased. The second reason is the fact that as
we become more sensitive to thermal motion of atoms, the mean energy loss by elastic scattering is decreasing progressively. 
At last, the increase of K above 5~keV is coming from the parameter $t_0$ which value is set to minimize the variation of K
around its mean value. We have set here $t_0$, the time necessary to go from 14~MeV to 800~keV by elastic scattering, equal to 
0.45$\mu$s. \\

	\subsection{Flux measurements}
\label{flux_measure}	
The $^3$He energy spectrum gives a relative validation measurement of the neutron flux description in relation to the
simulation. The energy spectrum deduced from the $^{233}$U fission rate gives the absolute neutron flux in the
lead block where the (n,$\gamma$) measurement is made, using the neutron source calibration.\par

		\subsubsection{Measurements with $^3$He counter}
		
Figure~\ref{taux_reac_he} shows the $^3$He (n,p) reaction rate measured and simulated by using ENDF/B-VI and
JENDL3.2 data bases for lead cross sections. It has first been checked that the $^3$He cross sections given by
the three data bases (ENDF/B-VI, JEF2.2, JENDL3.2) are quite identical, the one finally used being
ENDF/B-VI.\\
Above 100~keV the experimental spectrum shows a counting rate much higher than the simulated one: in this
energy range, corresponding to the neutron pulse, the counter sees many events due to $\gamma$-rays and other
processes, not directly due to neutrons,, which are not described by the simulation. This energy range will not
be exploited here.\\
Below 50~keV the experimental reaction rate is particularly well described by the simulation using JENDL3.2
for lead. The use of ENDF/B-VI leads to an underestimation: this comes probably from the missing $^{204}$Pb
isotope which contributes for 1.4\% in natural lead, and from a different description in the unresolved
resonance region. It must also be noticed than below 1~keV ENDF diffusion cross section for natural lead is
1.5\% lower than the JENDL one.\\
A weak discrepancy between simulations and measurement is observed around 0.5~eV, which can be due to a lack
of knowledge of impurity proportions in the lead, as Cadmium for instance.\\ 

		\subsubsection{Absolute neutron flux}
\label{abs_nflux}
	 
\paragraph{Irradiations: 14~MeV neutron transport in the lead block and monitoring calibration\\}

Irradiation of Nickel foils brings two types of results. The (n,2n) reactions $^{58}$Ni having a threshold
around 13~MeV, Ni targets put at different places in the lead block are sensitive to neutrons coming almost directly
from the TiT target of GENEPI. The dimming of the source neutron flux can be observed. This reduction follows
a $1/x^{4.5}$ law, x being the distance to the neutron source. The simulation of the propagation
shows a good agreement with the measurement.\\
Thanks to the Nickel foils irradiation the number of neutrons emitted by the TiT target in 4$\pi$ has been
determined with an incertainty of 15\% which comes mainly from (n,2n) cross section. The neutron number in this experiment
 is equal to 1.7$\times$10$^6$ n/pulse and which permits to calibrate the neutron monitoring.\\

\paragraph{$^{233}$U fission measurements\\}

As explained above, below 50~keV the neutron flux is well simulated using JENDL3.2 data base for lead. We
investigate now the most appropriate data base to describe $^{233}$U fission cross section and the subsequent
flux. Figure \ref{taux_reac_u3} shows the experimental and simulated $^{233}$U fission rate using ENDF/B-VI
and JENDL3.2 for $^{233}$U, that is to say $N_{U3}\varepsilon_f\sigma_f(E)\Phi(E)fN_{n/pulse}$, where N$_{U3}$ is the
number of $^{233}$U atoms of the target, $\varepsilon_f$ a factor including the detection efficiency and the
solid angle, $\sigma_f(E)$ the fission cross section, $\Phi(E)$ the neutron flux, f the GENEPI repetition rate
and $N_{n/pulse}$ the number of source neutron per pulse. 
The absolute fission rate per source neutron $N_{U3}\varepsilon_f\sigma_f(E)\Phi(E)$ can be determined by the 
knowledge of $N_{U3}\varepsilon_f$, experimentally obtained from the $\alpha$ activity $A_\alpha$ measurement of the target in the
same geometrical conditions: we obtain then $A_\alpha=N_{U3}\varepsilon_\alpha\lambda_{U3}$ with
$\varepsilon_f=\varepsilon_\alpha$ as the set-up is exactly the same for fission as for $\alpha$ detection,
and as the $^{233}$U target is very thin. The absolute fission rate per source neutron can then be written 
$N_f(E)=\frac{A_\alpha}{\lambda_{U3}}\sigma_f(E)\Phi(E)$ and we can finally deduce 

$$\sigma_f(E)\Phi(E)=\frac{N_f(E)\ln 2}{A_\alpha T_{1/2U3}fN_{n/pulse}}.$$\\

This leads to an absolute flux measurement between 1~eV and 40~keV per source neutron which can be compared to simulations. 
Despite the fact that the counting stastitics is a little bit too low, the reaction rate shape is well
described from 0.03~eV up to 100~eV within 10\%. A discrepancy can be seen between 100 and 500~eV for both
data bases, whereas between 2~keV and 10~keV the agreement is better with JENDL3.2.\\
Figure \ref{flux_u3} shows the neutron flux in the $^{233}$U target: experimental fission rate has been divided by the 
ENDF cross section widened by the lead block resolution, and is compared with simulated fission rates divided by both
ENDF and JENDL widened cross sections. This procedure sets the flux free from the data base defects. The flux is found
then well described from 0.03~eV to 100~keV. Around 2-3~eV a flux deficit is observed for the three plots which is due to an
insufficient broadening of the 1.8~eV resonance of $^{233}$U fission cross section. The general isolethargic behaviour of
neutron flux in the lead, i.e. $\Phi(E)\propto 1/E$, is also found. The slow variation of the flux above 1~eV is due to
neutron captures in the lead whereas below 1~eV the slope is stronger because of neutron escape from the block.\\

		\subsubsection{Capture rate in Au}
\label{capture_au}
As Gold capture cross section is well known and data bases are in good agreement, this element can be chosen as a
reference. 
In figure \ref{capt_au} the measured and normalized capture rate of a 125~$\mu$m Au foil is compared to simulations 
made again with both ENDF/B-VI and JENDL3.2 data bases for lead. Below 2~keV, as seen for (n,p) reaction on $^3$He, the
capture rate is described with ENDF/B-VI as well as with JENDL3.2. Above 2~keV the agreement is excellent with JENDL3.2
but the use of ENDF/B-VI brings a discrepancy of 20\% on the Au capture rate. The weak discrepancy between experiment
and the JENDL3.2 simulation in the 3-5~keV energy region will be discussed in \S~\ref{application}.\\

	\subsection{Conclusions about the flux}

We have seen in this section the high sensibility of the simulation on the choice of the data base for a material as
important as lead. A very good description of the neutron flux is obtained in the energy range 0.05~eV to 50~keV by
using JENDL3.2 for Lead for which we recommend the use of this data base.  
This is based in two independent measurements $^{3}$He(n,p) and $^{197}$Au(n,$\gamma$).\\
The results obtained by irradiation of Nickel foils and exploiting the 13~MeV-threshold reaction give the absolute
number of neutrons emitted by the TiT source and the source neutron propagation in the lead spectrometer. GENEPI neutron
production being necessary to know the absolute neutron flux in the lead block, we have shown the validity of this quantity by
the comparison of the experimental  $^{233}$U fission rate to the simulated one.\\
We can then conclude that between 14~MeV and 40~keV, despite the lack of measurement, the neutron transport is well
reproduced by the simulation using the JENDL3.2 data base. The neutron flux for energies lower than 40~keV measured with 3
different experimental method is also very well described by simulation within 5\%.\\

\section{Capture results}
\label{section4}
	\subsection{Thin an thick Au targets for simulation validation}
\label{au_validation}

In section~\ref{capture_au} we have presented results for a 125~$\mu$m thick Au foils as a mean of validation of the
neutron flux simulaton with JENDL3.2 for lead. All the simulations that will be shown in this section will use this data
base for neutron transportation in Lead. \\
As for Gold no dependance is seen on the data base, we can concentrate on the MCNP simulation ability to take into
account the self-shielding effects in the target. Three capture rate measurements have been made with Au targets of
different thicknesses~: 125~$\mu$m, 500~$\mu$m and 1250~$\mu$m. Figure~\ref{epaisseur_au} shows measurements and
simulations normalized in the resolved  resonance region, i.e. between 1~eV and 100~eV. This region has been chosen as
the cross section of the studied nuclide is well known. A very good agreement is found over a wide energy range (0.05~eV
up to 30~keV) for three thicknesses. The two first resonances of Au at 4.9~eV and 60~eV are clearly seen. The reduction
of the capture rate maximum and the broadening of the 4.9~eV resonance is well taken into account by simulations. 
Considering the energy resolution of the lead spectrometer, it is not possible to separate high energy resonances. But
calculations reproduce rather well the variations of the cross section.\\
The only discrepancy greater than 10\%  is found between 2~keV and 5~keV for all thicknesses. So, it is not due to a bad
description of the self-shielding effects and cannot be attributed to a wrong neutron flux either (cf.
section~\ref{flux_measure}).
This discrepancy has the same magnitude for the three thicknesses.\\
This behaviour can be understood as follows. The total cross section $\sigma_{tot} = \sigma_{el}+\sigma_{(n,\gamma)}$ is
correct. But $\sigma_{(n,\gamma)}$ has been underestimated while $\sigma{el}$ is overestimated by the same quantity. Since the
self-shielding effect is mainly controled by $\sigma_{tot}$ (which is supposed constant), although in $\sigma_{(n,\gamma)}$ has
resonances are present, the underestimation of (n,$\gamma$) reaction does not depend on target thickness.

Considering these results we propose a so called "correction table" for the capture cross section in this energy range
(i.e. when the discrepancy between simulation and measurement is more than 5\%). These correction factors are resumed in
table~\ref{correc_au}.\\

	\subsection{Experimental and simulation data on (n,$\gamma$) reaction rates on various samples}
\label{application}
		\subsubsection{Tantalum}
Measurements have been made for three thicknesses (100~$\mu$m, 200~$\mu$m and 2000~$\mu$m). Data bases available for
this element are ENDF/B-VI, JEF2.2 and  JENDL3.2. All the results are presented in figure~\ref{epaisseur_ta}. First of all,
the good agreement between experiment and simulation made with the different data bases can be noticed from 0.1~eV to 
300~eV, whatever the thickness is. Self-shielding effects in Tantalum are well taken into account. Above 300~eV for the
thinest targets the agreement is better with JENDL3.2. The maximum discrepancy with JEF2.2 and ENDF/B-VI is 15\% at
maximum.\\
In the case of the thickest (2000~$\mu$m) target, all data bases are in good agreement between 0.1~eV and 300~eV. In the
300~eV-3~keV energy range, a discrepancy of about 40\% is found for ENDF/B-VI and JEF2.2 simulations in comparison with
experiment while 
a good agreement is found for JENDL3.2. It must be noticed that JENDL3.2 gives resonances up to 2.2~keV whereas the
other ones are smooth above 300~eV. 
Self-shielding effects are well taken into account if we use this data base because of the good resonance description of the
cross section.\\
Above 3~keV the reaction rate is better described by ENDF. In this energy range self-shielding can be neglected. The 
discrepancy with JENDL  here is not due to the cross section structure but to an absolute underestimation of it.\\
We can conclude that in general ENDF and JEF data bases give simulation results in worst agreement with measurements than
JENDL and then advise users to take JENDL for Tantalum between 0.1~eV and 30~keV. However we can propose as for Gold a
correction table for the JENDL3.2 $^{181}$Ta capture cross section above 3~keV (see table~\ref{correc_ta}).\\

		\subsubsection{Indium}

In figure~\ref{epaisseur_in} experimental and simulated reaction rates of Indium are presented, for the three data bases and
 for several thicknesses: 300~$\mu$m, 500~$\mu$m and 2000~$\mu$m. All data bases show an excellent agreement with experiment
from 0.1~eV to 1~keV for the three thicknesses. Between 1~keV and 2~keV, simulations using ENDF/B-VI and JENDL3.2
underestimate the real capture rate. JEF2.2 shows an agreement within 10\%. Above 2~keV, In capture cross section given by   
ENDF/B-VI or JENDL3.2 is 25\% higher than the one given by JEF2.2, but the three cross sections do not present resonances in this energy
range. The strongest discrepancy is obtained with JENDL3.2 which shows a 30\% overestimation around 3~keV. ENDF/B-VI shows
also varying discrepancies in this region with a maximum of 25\% at 500~eV. JEF2.2 shows a mean underestimation of the
capture rate in this energy range of about 20\%. For this element we notice that no data base gives a satisfactory results above
2~keV. Then we propose a correction table to be applied to JEF2.2 in table~\ref{correc_in}.\\

		\subsubsection{Thorium}

Figure~\ref{epaisseur_th} presents calculated and measured capture  rates for three Thorium samples. First of all we observe
a good agreement within 5\% between ENDF and JEF simulations and experiment for the three thicknesses (175~$\mu$m,
1000~$\mu$m, and 4000~$\mu$m) in the 12~eV-40~keV energy range. JENDL data base shows a discrepancy with experiment of about 15\% between
600~eV and 5~keV. In this range, the capture cross sections of the three data bases are described by a resonance structure,
but the mean level of the JENDL one is about 15\% weaker than the ENDF and JEF ones.\\
Below 10~eV, considering the weak value of the Th capture cross section ($\sim$0.3~barn), the quick decrease of the neutron
flux and the strong background observed due to Thorium radioactivity, the signal to background ratio in this energy range
is weak. This explains the strong fluctuations in the experimental spectra. In this region it is impossible to say which data
base is the most satisfactory despite the discrepancies observed between ENDF/B-VI and JENDL3.2 in the 1~eV to 10~eV
region.\\
Consequently for $^{232}$Th ENDF/B-VI and JEF2.2 can be chosen as they give the best agreement with experiment. In this case, 
because of relatively large errors we do not propose a correction table as for the previous nuclides.\\

		\subsubsection{Silver}

Results for simulated and measured reaction rates for three thicknesses are resumed in figure~\ref{epaisseur_ag}. We observe
a very good agreement between experiment and simulation and also a good taking into account of self-shielding effects in the
0.1~eV-600~eV energy range.\\
ENDF data base shows an important discrepancy of about 35\% between 600~eV and 3~keV. In our case, for ENDF, natural Silver
is composed with two isotopes: $^{107}$Ag and $^{109}$Ag. The discrepancy is reduced to 10\% above 3~keV where the capture
cross section is described by a smoothed curve.\\
JEF2.2 and JENDL3.2 data bases present an underestimation of the reaction rate of about 20\% between 600~eV and 1~keV.
JENDL3.2 is in agreement with experiment for the three thicknesses, from 1.2~keV to 4~keV. Reaction rates given by JEF2.2
show an overestimation in the 1~keV-4~keV range of about 20\%. But capture rate is well described by this base above 4~keV.\\
Consequently if we can dismiss ENDF/B-VI data base because of the too important discrepancy observed above 600~eV, our choice
is settled on JEF2.2 or JENDL3.2. We propose a correction table (table~\ref{correc_ag}) to be applied to JENDL3.2 because of
its description with resonances up to 7~keV which seems to have more sensitivity on self-shielding effects as it was
shown for the case of Tantalum.\\

		\subsubsection{Technetium}

We also did a measurement with Technetium with a sample of 344~mg of $^{99}$Tc diluted in 6.4~g of Aluminium. Simulations
have been performed  with the three data bases. Results are shown in figure~\ref{reac_tc}.\\
A good agreement is obtained in the 0.1~eV-40~eV energy range between simulations and experiment. The position of the 20~eV
resonance is better described by JENDL3.2 and JEF2.2 but with ENDF it is shifted to about 21~eV: despite the weak resolution
of the lead block we are sensitive to small shifts.\\
Between 40~eV and 4~keV none of the three data bases give a better description of the capture rate than 10\%. ENDF/B-VI is
the worst one with a discrepancy of 35\% from 80~eV to 500~eV. The rate simulated with JENDL3.2 shows also a 25\%
underestimation between 80~eV and 3~keV. The data base which fits the best the experiment is JEF2.2. We find a discrepancy of
20\% between 150~eV and 700~eV which appears like an overestimation of the reaction rate in comparison with the experiment.
But we can also  observe that resonance positions given by experiment - 165~eV and 350~eV - are better described by JENDL3.2
than by JEF2.2, this last one giving respectively 180~eV and 365~eV. Above 1~keV, despite observed experimental errors,
JEF2.2 describes more satisfactory the experimental reaction rate than ENDF and JENDL.\\
Therefore ENDF shows the worst agreement with our experiment. Even if the resonance structure is similar in JEF and JENDL,
this last one gives a better general description of the reaction rate over all the energy range studied (0.1~eV-40~keV). We
will thus apply the correction table~\ref{correc_tc} to JENDL cross section.\\

{\em mentionner les resultats de Lepretre}\\

		\subsubsection{Manganese}

Results of simulated and measured capture rate for $^{55}$Mn are shown in figure~\ref{reac_mn}. The only energy range to be
in agreement is the 0.1~eV to 40~eV region. Above 40~eV strong discrepancies appear between the different data bases and
experiment. We can notice that resonance positions given by the different capture cross section are not satisfactory. The
width of the 336~eV resonance is not well described by ENDF/B-VI and JENDL3.2. Experiment presents a maximum capture rate at
2.2~keV which is not reproduced by simulations.\\
In that case, considering the strong discrepancies observed between simulated capture rates and our measurement, we cannot
propose a correction table, but we suggest that the capture cross section of this nuclide should be re-measured and re-evaluated.\\

		\subsubsection{Naturel Nickel}

A measurement with a 3~mm natural Nickel sample has been done. This element is of interest as it is present in steel composing structure materials
of nuclear reactors. Experiment and simulations are shown in figure~\ref{reac_ni}.\\
A good agreement is found from 0.1~eV to 50~eV. But strong discrepancies are observed on reaction rates given by ENDF/B-VI and
JENDL3.2 in the 1~keV-10~keV range. JENDL3.2 gives a maximum reaction rate at 1350~eV which does not appear with ENDF/B-VI or
JEF2.2. Cross
section structures are very different above 1~keV. If we compare this simulation to experiment, we see that the simulation
is in desagreement with experimental data over all this energy range. A capture rate maximum is observed at 16~keV in simulations, whereas it is at
12.5~keV for experiment.\\
As in the case of Mn, we do not propose a correction table because of the too big discrepancies. A new measurement and
evaluation of the Nickel capture cross section should be made.\\

\section*{Conclusion}

Using precise measurements in a LSDS and their very detailed numerical simulation with MCNP code, it
has been possible to check the validity of different data bases.\\
The first result was to show the sensitivity (1\%) of slowing down time to the elastic scattering of the Lead medium. In Lead
medium, precise energy dependence of the neutron flux allows us to strongly recommend the JENDL3.2 data base for Lead. This
recommendation is also fully supported by (n,$\gamma$) reaction rate on Au targets.\\

The second result is that the presented methodology (measurement, data analysis, simulation) provides a check of the validity
of the data base within a 5\% precision. As foreseen it appears that for neutron energy below 200~eV the (n,$\gamma$) cross
section are well determined. Usually the different data bases do not differ significantly.

At higher energy the (n,$\gamma$) cross sections from data bases are quite different. The experimental data provide either a
recommendation for using a special data base and a correction factor table in order to reprodure experimental data with an
error of less than 10\%. In some cases (Ni and Mn) the recommendation is to re-measure the (n,$\gamma$) cross section. \\
The
overall conclusion is that despite the poor energy resolution of the LSDS, precise indications on the validity or deviations
of data bases cross section can be obtained either for elastic and capture cross section in Lead or small samples 
(n,$\gamma$) cross sections, and with rather short beam time (about a day). It must be pointed out that measurements with different thicknesses show the importance of
resonance in the so-called unresolved resonance region. Finally the LSDS appears a very efficient tool to test different
elements cross sections and recommend a particular data base and in some case the use of correction factors.\\

\section*{Acknowledgement}

The authors would like to thank the neutron generator builders and operator teams, the electronic development group for
developping an efficient data acquisition system. 
This work has been partly supported by GEDEON, concerted research
organisations, and PACE CNRS program.\\

\newpage

\begin{table}[hbtp]
\begin{center}
\begin{tabular}[t]{|c|c|c|}
	\hline
	Time ($\mu$s)	& 	Energy (eV)	&	Element	\\
	\hline
	178		&	5.3		&	Ag	\\
	\hline
	50		&	67		&	Ce	\\
	\hline
	34.5		&	140		&	Ce+Co	\\
	\hline
	26		&	244		&	Ce	\\
	\hline
	16.3		&	612		&	Cu	\\
	\hline
	9.3		&	1.82		&	Pb+Ce	\\
	\hline
	6.6		&	3.5		&	Pb	\\
	\hline
\end{tabular}
\end{center}
\caption{Main identified resonances of the background with their associated element.}
\label{bdf1}
\end{table}

\begin{table}[hbtp]
\begin{center}
\begin{tabular}[t]{|c|c|c|c|c|c|}
	\hline
  Energy Range & $\langle \sigma_{(n,\gamma)}(E) \rangle$ & \multicolumn{3}{c|}{$\frac{\Delta\sigma\phi_{(C-E)}}{\sigma\phi_{E}}$}	&  Correction  \\
\cline{3-5}
  (eV)         &	(barns)		       & 		e=125$\mu$m			   &   		
  e=500$\mu$m   		 &   			e=1250$\mu$m		       &  factor  \\
	\hline
  725.70$<$E$<$916.27  & 10.36  &  -1.11e-02     &	  -1.32e-02   &    -1.68e-02	&   1.00       \\
 916.27$<$E$<$1156.89  &  7.35  &  -4.37e-02     &	  -3.99e-02   &    -8.74e-02	&   1.06       \\
1156.89$<$E$<$1460.68  &  5.55  &  -4.81e-02     &	  -5.69e-02   &    -5.22e-02	&   1.05       \\
1460.68$<$E$<$1844.26  &  4.10  &  -3.89e-02     &	  -8.57e-02   &    -7.42e-02	&   1.07       \\
1844.26$<$E$<$2328.56  &  2.64  &  -1.12e-01     &	  -1.28e-01   &    -1.25e-01	&   1.12       \\
2328.56$<$E$<$2940.04  &  2.39  &  -1.43e-01     &	  -1.61e-01   &    -1.76e-01	&   1.16       \\
2940.04$<$E$<$3712.10  &  1.97  &  -1.47e-01     &	  -1.25e-01   &    -1.35e-01	&   1.14       \\
3712.10$<$E$<$4686.89  &  1.99  &  -1.27e-01     &	  -1.00e-01   &    -8.68e-02	&   1.11       \\
4686.89$<$E$<$5917.67  &  1.89  &  -3.67e-03     &	  -1.38e-02   &     2.88e-02	&   1.00       \\
5917.67$<$E$<$7471.65  &  1.57  &   8.52e-03     &	   3.39e-02   &     5.62e-02	&   1.00       \\
7471.65$<$E$<$9433.71  &  1.33  &   1.93e-02     &	   2.33e-02   &     4.59e-02	&   1.00       \\
	\hline
\end{tabular}
\end{center}
\caption{Proposition of correction factors for ENDF/B-VI capture cross section of $^{197}$Au in the 1~keV-5~keV energy
range. The data base mean cross section, the relative difference between calculated (C) and experimental (E) $\sigma\Phi$ 
for different thicknesses, and the final correction factor are given.}
\label{correc_au}
\end{table}

\newpage

\begin{table}[hbtp]
\begin{center}
\begin{tabular}[t]{|c|c|c|c|c|c|}
	\hline
  Energy Range & $\langle \sigma_{(n,\gamma)}(E) \rangle$  & \multicolumn{3}{c|}{$\frac{\Delta\sigma\phi_{(C-E)}}{\sigma\phi_{E}}$}	&  Correction  \\
\cline{3-5}
  (eV)         &	(barns)		       & 		e=100$\mu$m			   &   		
  e=200$\mu$m   		 &   			e=2000$\mu$m		       &  factor  \\
	\hline
    725.70$<$E$<$916.27  &  9.30  &  -1.94e-02     &	-3.53e-02   &	 -7.36e-02    &   1.00       \\
   916.27$<$E$<$1156.89  &  9.68  &  -3.61e-02     &	-1.15e-01   &	 -9.60e-02    &   1.10       \\
  1156.89$<$E$<$1460.68  &  6.86  &  -7.64e-02     &	-1.31e-01   &	 -1.31e-01    &   1.13       \\
  1460.68$<$E$<$1844.26  &  5.57  &  -5.13e-02     &	-1.31e-01   &	 -1.70e-01    &   1.17       \\
  1844.26$<$E$<$2328.56  &  3.71  &  -9.37e-02     &	-1.47e-01   &	 -1.46e-01    &   1.15       \\
  2328.56$<$E$<$2940.04  &  3.89  &  -7.56e-02     &	-8.38e-02   &	 -1.21e-01    &   1.12       \\
  2940.04$<$E$<$3712.10  &  3.32  &  -3.54e-02     &	-7.72e-02   &	 -1.01e-01    &   1.10       \\
  3712.10$<$E$<$4686.89  &  2.86  &  -4.87e-02     &	-1.13e-01   &	 -1.03e-01    &   1.10       \\
  4686.89$<$E$<$5917.67  &  2.43  &  -2.62e-02     &	-8.86e-02   &	 -1.18e-01    &   1.12       \\
  5917.67$<$E$<$7471.65  &  2.04  &  -5.72e-02     &	-1.17e-01   &	 -1.51e-01    &   1.15       \\
  7471.65$<$E$<$9433.71  &  1.71  &  -2.33e-02     &	-1.40e-01   &	 -1.67e-01    &   1.17       \\
 9433.71$<$E$<$11911.00  &  1.36  &  -5.24e-02     &	-1.33e-01   &	 -1.84e-01    &   1.18       \\
11911.00$<$E$<$15038.84  &  1.15  &  -6.62e-02     &	-1.48e-01   &	 -2.03e-01    &   1.20       \\
15038.84$<$E$<$18988.04  &  0.98  &  -8.65e-02     &	-1.90e-01   &	 -2.14e-01    &   1.21       \\
	\hline
\end{tabular}
\end{center}
\caption{Proposition of correction factors for JENDL3.2 capture cross section of $^{181}$Ta in the 800~eV-20~keV energy
range. The data base mean cross section, the relative difference between calculated (C) and experimental (E) $\sigma\Phi$ 
for different thicknesses, and the final correction factor are given.}
\label{correc_ta}
\end{table}

\newpage

\begin{table}[hbtp]
\begin{center}
\begin{tabular}[t]{|c|c|c|c|c|c|}
	\hline
  Energy Range & $\langle \sigma_{(n,\gamma)}(E) \rangle$  & \multicolumn{3}{c|}{$\frac{\Delta\sigma\phi_{(C-E)}}{\sigma\phi_{E}}$}	&  Correction  \\
\cline{3-5}
  (eV)         &	(barns)		       & 		e=300$\mu$m			   &   		
  e=500$\mu$m   		 &   			e=2000$\mu$m		       &  factor  \\
	\hline
  1156.89$<$E$<$1460.68  &  1.92 &  1.69e-01 & -5.33e-02 & -6.70e-02  &  1.00	\\
  1460.68$<$E$<$1844.26  &  1.74 & -8.78e-02 & -1.28e-01 & -1.51e-01  &  1.14	\\
  1844.26$<$E$<$2328.56  &  1.58 & -1.29e-01 & -1.65e-01 & -1.60e-01  &  1.16	\\
  2328.56$<$E$<$2940.04  &  1.44 &  8.41e-02 & -1.45e-01 & -1.52e-01  &  1.15	\\
  2940.04$<$E$<$3712.10  &  1.33 & -9.02e-02 & -1.31e-01 & -1.36e-01  &  1.13	\\
  3712.10$<$E$<$4686.89  &  1.24 & -1.86e-01 & -1.69e-01 & -1.27e-01  &  1.15	\\
  4686.89$<$E$<$5917.67  &  1.15 & -2.01e-01 & -1.77e-01 & -1.16e-01  &  1.15	\\
  5917.67$<$E$<$7471.65  &  1.07 & -1.95e-01 & -1.91e-01 & -1.31e-01  &  1.16	\\
  7471.65$<$E$<$9433.71  &  1.00 & -2.28e-01 & -1.85e-01 & -1.49e-01  &  1.17	\\
 9433.71$<$E$<$11911.00  &  0.91 & -2.00e-01 & -1.68e-01 & -1.45e-01  &  1.16	\\
11911.00$<$E$<$15038.84  &  0.84 & -1.95e-01 & -2.07e-01 & -1.62e-01  &  1.18	\\
15038.84$<$E$<$18988.04  &  0.77 & -2.12e-01 & -1.86e-01 & -1.76e-01  &  1.18	\\
18988.04$<$E$<$23974.29  &  0.70 & -1.90e-01 & -1.94e-01 & -1.87e-01  &  1.19	\\
23974.29$<$E$<$30269.95  &  0.62 & -2.10e-01 & -2.24e-01 & -1.94e-01  &  1.21	\\
	\hline
\end{tabular}
\end{center}
\caption{Proposition of correction factors for JEF2.2 capture cross section of natural Indium in the 1.5~keV-40~keV energy
range. The data base mean cross section, the relative difference between calculated (C) and experimental (E) $\sigma\Phi$ 
for different thicknesses, and the final correction factor are given.}
\label{correc_in}
\end{table}

\newpage

\begin{table}[hbtp]
\begin{center}
\begin{tabular}[t]{|c|c|c|c|c|c|}
	\hline
  Energy Range & $\langle \sigma_{(n,\gamma)}(E) \rangle$  & \multicolumn{3}{c|}{$\frac{\Delta\sigma\phi_{(C-E)}}{\sigma\phi_{E}}$}	&  Correction  \\
\cline{3-5}
  (eV)         &	(barns)		       & 		e=100$\mu$m			   &   		
  e=300$\mu$m   		 &   			e=1200$\mu$m		       &  factor  \\
	\hline
    600.00$<$E$<$725.70  &  3.57 & -2.55e-01 & -2.82e-01 & -2.83e-01  &  1.27  \\
    725.70$<$E$<$916.27  &  2.50 & -1.91e-01 & -2.68e-01 & -2.27e-01  &  1.23  \\
   916.27$<$E$<$1156.89  &  3.01 &  3.53e-02 & -8.61e-02 &  2.05e-02  &  1.00  \\
  1156.89$<$E$<$1460.68  &  3.54 &  1.96e-01 & -1.09e-02 &  9.71e-02  &  0.91  \\
  1460.68$<$E$<$1844.26  &  3.04 &  4.56e-02 & -4.74e-02 &  1.46e-02  &  1.00  \\
  1844.26$<$E$<$2328.56  &  2.57 &  9.66e-03 & -1.32e-01 & -3.70e-02  &  1.05  \\
  2328.56$<$E$<$2940.04  &  2.05 &  1.40e-03 & -9.08e-02 & -6.95e-02  &  1.05  \\
  2940.04$<$E$<$3712.10  &  1.99 & -6.10e-02 & -1.19e-01 & -8.17e-02  &  1.09  \\
  3712.10$<$E$<$4686.89  &  1.76 & -1.04e-01 & -1.38e-01 & -9.06e-02  &  1.11  \\
  4686.89$<$E$<$5917.67  &  1.66 & -8.09e-02 & -1.56e-01 & -8.05e-02  &  1.11  \\
  5917.67$<$E$<$7471.65  &  1.48 & -6.96e-02 & -1.03e-01 & -7.16e-02  &  1.08  \\
  7471.65$<$E$<$9433.71  &  1.52 & -8.51e-02 & -9.15e-02 & -6.76e-02  &  1.08  \\
 9433.71$<$E$<$11911.00  &  1.33 & -7.08e-02 & -9.93e-02 & -7.65e-02  &  1.08  \\
11911.00$<$E$<$15038.84  &  1.23 & -5.14e-02 & -1.12e-01 & -8.46e-02  &  1.08  \\
15038.84$<$E$<$18988.04  &  1.13 & -9.65e-02 & -1.40e-01 & -9.62e-02  &  1.11  \\
18988.04$<$E$<$23974.29  &  1.03 & -2.64e-02 & -1.42e-01 & -9.39e-02  &  1.09  \\
23974.29$<$E$<$30269.95  &  0.92 & -6.29e-02 & -1.51e-01 & -9.85e-02  &  1.10  \\
	\hline
\end{tabular}
\end{center}
\caption{Proposition of correction factors for JENDL3.2 capture cross section of natural Silver in the 600~eV-40~keV energy
range. The data base mean cross section, the relative difference between calculated (C) and experimental (E) $\sigma\Phi$ 
for different thicknesses, and the final correction factor are given.}
\label{correc_ag}
\end{table}

\newpage

\begin{table}[hbtp]
\begin{center}
\begin{tabular}[t]{|c|c|c|c|}
	\hline
  Energy Range & $\langle \sigma_{(n,\gamma)}(E) \rangle$  & $\frac{\Delta\sigma\phi_{(C-E)}}{\sigma\phi_{E}}$	&  Correction  \\
  (eV)         &	(barns)		       &	m=344mg						&	factor	\\
	\hline
   70.49$<$E$<$ 89.00  &  1.16 & -3.11e-01  &  1.31	  \\
   89.00$<$E$<$112.37  &  3.99 & -2.54e-01  &  1.25	  \\
  112.37$<$E$<$141.87  &  5.04 & -2.85e-01  &  1.28	  \\
  141.87$<$E$<$179.13  & 12.35 & -2.84e-01  &  1.28	  \\
  179.13$<$E$<$226.17  &  7.21 & -2.69e-01  &  1.27	  \\
  226.17$<$E$<$285.56  &  4.88 & -3.27e-01  &  1.33	  \\
  285.56$<$E$<$360.55  &  6.09 & -2.52e-01  &  1.25	  \\
  360.55$<$E$<$455.23  &  4.83 & -1.33e-01  &  1.13	  \\
  455.23$<$E$<$574.77  &  3.62 & -1.62e-01  &  1.16	  \\
  574.77$<$E$<$725.70  &  3.36 & -1.69e-01  &  1.17	  \\
  725.70$<$E$<$916.27  &  3.22 & -2.24e-01  &  1.22	  \\
 916.27$<$E$<$1156.89  &  2.73 & -2.16e-01  &  1.22	  \\
1156.89$<$E$<$1460.68  &  2.81 & -2.88e-01  &  1.29	  \\
1460.68$<$E$<$1844.26  &  1.78 & -2.53e-01  &  1.25	  \\
1844.26$<$E$<$2328.56  &  1.81 & -2.49e-01  &  1.25	  \\
2328.56$<$E$<$2940.04  &  1.87 & -1.07e-01  &  1.11	  \\
2940.04$<$E$<$3712.10  &  2.33 & -4.59e-02  &  1.00	  \\
	\hline
\end{tabular}
\end{center}
\caption{Proposition of correction factors for JENDL3.2 capture cross section of $^{99}$Tc in the 80~eV-4~keV energy
range. The data base mean cross section, the relative difference between calculated (C) and experimental (E) $\sigma\Phi$ 
for a 344~mg target, and the final correction factor are given.}
\label{correc_tc}
\end{table}


\end{document}